\documentclass[10pt,journal,compsoc]{IEEEtran}

\ifCLASSOPTIONcompsoc
  \usepackage[nocompress]{cite}
\else
  \usepackage{cite}
\fi
\ifCLASSINFOpdf
\else
\fi
\usepackage{graphicx}
\usepackage{float}
\usepackage{placeins}
\usepackage{multirow}
\usepackage{hyperref}
\usepackage{amsmath}
\usepackage[table,xcdraw]{xcolor}
\usepackage{booktabs}
\usepackage{subcaption}
\addtocounter{figure}{-1}

\begin{document}
%

\title{Semi-supervised Learning for Identifying the Likelihood of Agitation in People with Dementia}
%
%

\author{Roonak Rezvani,
        Samaneh Kouchaki,
        Ramin Nilforooshan,
        David J. Sharp, 
        \protect\\ and Payam Barnaghi, Senior Member, IEEE 
\IEEEcompsocitemizethanks{\IEEEcompsocthanksitem R. Rezvani and S. Kouchaki are with the Centre for Vision, Speech and Signal Processing (CVSSP), University of Surrey.\protect\\
\IEEEcompsocthanksitem R. Nilforooshan is with Surrey and Borders Partnership NHS Foundation Trust. \protect\\
\IEEEcompsocthanksitem D. J. Sharp and P. Barnaghi are with Imperial College London. \protect\\
\IEEEcompsocthanksitem S. Kouchaki, R. Nilforooshan, D. Sharp and P. Barnaghi are with UK Dementia Research Institute Care Research and Technology Centre, Imperial College London and the University of Surrey, UK.
}
}




\IEEEtitleabstractindextext{%
\begin{abstract}
Interpreting the environmental, behavioural and psychological data from in-home sensory observations and measurements can provide valuable insights into the health and well-being of individuals. Presents of neuropsychiatric and psychological symptoms in people with dementia have a significant impact on their well-being and disease prognosis. Agitation in people with dementia can be due to many reasons such as pain or discomfort, medical reasons such as side effects of a medicine, communication problems and environment. This paper discusses a model for analysing the risk of agitation in people with dementia and how in-home monitoring data can support them. We proposed a semi-supervised model which combines a self-supervised learning model and a Bayesian ensemble classification. We train and test the proposed model on a dataset from a clinical study. The dataset was collected from sensors deployed in $96$ homes of patients with dementia. The proposed model outperforms the state-of-the-art models in recall and f1-score values by $20\%$. The model also indicates better generalisability compared to the baseline models.
\end{abstract}

\begin{IEEEkeywords}
Dementia care, agitation, semi-supervised, partially labelled, imbalanced dataset.
\end{IEEEkeywords}}

\maketitle

\IEEEdisplaynontitleabstractindextext

%
\IEEEpeerreviewmaketitle

\IEEEraisesectionheading{\section{Introduction}\label{sec:introduction}}
\IEEEPARstart{C}{ontinuous} monitoring enables us to support older adults and people with progressive and chronic illnesses in managing their daily life. Wearable and smart sensors have been parts of people's homes and lives that can be employed for monitoring purposes. Dementia is a neurodegenerative disease with no disease modifying treatment. Dementia presents with global cognitive decline and many physical and psychological sings and symptoms. The behavioural and physical symptoms of dementia include agitation, irritability, aggression, memory loss, decision-making problems, lack of awareness and Urinary Tract Infection (UTI) \cite{nhs:2017}. The data collected from in-home sensors could support people with dementia's independent living. Sensors that do not intervene in people's lives, such as Passive Infra-Red (PIR), motion and door sensors, are beneficial. The data from sensors also provide clinicians with information regarding patients' daily activities. However, the data are subject to missing values and are irregular. Moreover, the sensors are deployed in uncontrolled homes and environments, so the data are noisy. 

As part of the research in the Care Research and Technology Centre at the UK Dementia Research Institute (UK DRI), we have been developing and deploying in-home monitoring technologies to support people affected by dementia. Our research has led to developing a digital platform that allows collecting and integrating in-home observations and measurements using network-connected sensory devices. This paper discusses how in-home monitoring data and machine learning algorithms are used to detect the risk of agitation in people with dementia.

Machine Learning algorithms can be applied to detect any change of behaviour and symptoms from the collected sensory data. The information can then be used to inform clinicians to take further actions and help patients. However, to have a clinically-useful model, advanced machine learning techniques usually require large labelled data, which is difficult to collect and needs many manual observations that are not easily feasible in the current healthcare setting. Our dataset does not contain huge amounts of agitation labels, and the labelled data are imbalanced as most labels are from non-agitation episodes. Moreover, the data patterns are related to the home, as the sensory data were collected from multi-occupied properties. However, we are trying to link these patterns to individual patients. The goal of the paper is to deal with these limitations and provide a generalisable model. In this paper, we introduce a semi-supervised platform that uses the information from labelled and unlabelled data collected from sensors in peoples with dementia's homes to detect agitation risk. 

\section{Related Work}

Internet of Things and smart home technologies help healthcare systems worldwide as they provide an opportunity to monitor people in their homes without any disruption to their everyday activities \cite{smarthome}. The in-home sensory observations and measurements can detect different activities, anomalies, and changes to vital signals \cite{smartelderly}. There are also research studies on possible opportunities for utilising the sensors and imaging technologies in detecting neurological diseases such as dementia which is cost-effective \cite{biomarkerdementia}. Most of the existing studies on using sensory data in this area have focused on detecting and analysing daily living activities. In some cases, researchers have applied model-based clustering and also Naive Bayesian classification \cite{tapia2010agents} to detect daily living activities. Hidden Markov Models (HMMs) \cite{hmm}, Probabilistic models \cite{probabilistic}, and Recurrent Neural Networks (RNNs) \cite{rnn1, rnn2} are widely adopted in this area. Other models such as K-means \cite{kmeans}, Random Forest (RF) \cite{randomforest1}, and rule-based methods \cite{rulebased} have also been used to analyse in-home sensory data.

According to the World Health Organisation (WHO) \cite{world2015ageing}, dementia is one of the health challenges, which has a significant impact on healthcare systems. Existing research shows the potential of employing in-home sensory devices in the future of clinical monitoring and assessments \cite{biomarkerdementia}. Assisting dementia patients in their home is one important application, which has been rarely addressed \cite{moyle2019promise}. Most of the current literature has focused on activity detection to monitor older adults to increase their ability to live independently in their homes. For example, \cite{urwyler2017cognitive} suggests detecting daily living activities with unobtrusive ambient sensors. The paper provides a comparison between healthy subjects and dementia patients and demonstrates a promising usage of these sensors in detecting the cognitive variations between different groups of subjects.

Capturing early symptoms and aiding in-home dementia care have been among the important research preventing unplanned and avoidable hospital admissions and improving the quality of life of people affected by dementia \cite{husebo2020corrigendum}. The characteristics of these behavioural symptoms (e.g., agitation) make it hard to choose suitable algorithms and sensory technologies. There are not many clinical tools to detect agitation or large dataset available to train machine learning algorithms \cite{yefimova2012using}. Moreover, there are variations in the daily trends of different patients. In addition, the privacy of patients impacts the possible sensory setting \cite{yefimova2012using}.

Semi-supervised learning can be used when there are limited labelled data and a large set of unlabelled data. Semi-supervised learning has shown to improve the model's performance with information from the unlabelled data \cite{semisurvey}. However, we need to consider if the number of unlabelled samples is sufficient to increase performance \cite{semi-book}. Various types of semi-supervised learning have been used in state-of-the-art algorithms. Unlabelled data can be integrated into learning by extracting their useful information with simple or Variational autoencoders \cite{autoencoder, variationalauto, semisurvey}. Pre-clustering the labelled and unlabelled data and training a classifier for each cluster can be considered as another approach \cite{pre-cluster}. However, it is challenging to train a classifier on clusters with insufficient labelled data. 

There are some state-of-the-art models which include unlabelled data directly into objective or optimisation functions. Mean teacher is in this category, which considers teacher and student models in learning \cite{meanteacher}. The student model carries the latest set of weights, and the teacher model has the averaged weights \cite{meanteacher}. Verma et al. proposed a Mixup model as a linear combination of data points and labels \cite{mixup}. The virtual adversarial training method assumes that adding a small amount of noise into the data points does not change the label of the data \cite{virtualat}. However, add more noise to noisy data can be a difficult decision.

The pseudo-labelling techniques can be used either during pre-processing or training steps. A classifier can generate pseudo-labels for the unlabelled data and use those labels in learning the final model \cite{pseudolabel}. It can also be treated as a self-training strategy. In each training iteration, self-training uses labelled data and the most confident pseudo-labels from the previous training iteration. However, choosing the best measure for confidence analysis can be challenging. Another technique that can be categorised as pseudo-labelling is self-supervised representation learning. In this technique, pseudo-labels are generated by transforming the unlabelled data to latent representations and training a classifier to recognise the different transformations. The goal here is to learn generalisable features from unlabelled data \cite{selfsupervised}.

Another challenge in the healthcare setting is learning from imbalanced data that happens when the distribution of samples is asymmetric. Therefore, the model is biased towards the majority class \cite{surveyimbalance} while the minority class is usually the most important. There are different approaches to overcome this issue. Oversampling the minority class or undersampling the majority class are among the state-of-the-art solutions \cite{surveyimbalance}. SMOTE is a widely used algorithm that generates synthetic data for minority class \cite{smote}. However, these models need a clear structure to generate high-quality samples. Also, removing samples in a situation where the number of samples is not enough can impact the performance. Another approach uses cost-sensitive models that assign a higher cost to minority class \cite{costsense}. However, choosing the proper value of cost to prevent learning a biased model is complex. 

Ensemble learning is another possible approach in this area. Combining the results from different classifiers with different data views can make the final results less biased \cite{ensemble}. However, the number of classifiers and the approach to combine results at the decision layer need to be set \cite{surveyimbalance}. One way of combining different classifiers is the Bayesian process. The Bayesian model generates a confusion matrix for each base classifier to evaluate their performance \cite{bayesian}. Bayesian Classifier Combination neural Network (BCCNet) is an extension to the Bayesian combination algorithm \cite{bccnet}. BCCNet combines the base classifiers using a neural network that uses a Bayesian combination algorithm in the training process.

This research proposes a semi-supervised process that can be used in the healthcare area and overcome the existing challenges. As a case study, we analyse the agitation risk using partially labelled data collected by unobtrusive environmental sensors. This research is part of a study to develop an integrated digital platform to support people affected by dementia \cite{enshaeifar2018internet}. In our previous research, we developed a hierarchical algorithm to detect agitation from sensory data \cite{enshaeifar2018health}. The model was used in a clinical study to collect and validate the results. Our previous algorithm's key issues were lack of access to extensive training data and generalisability \cite{enshaeifar2018health}. In this paper, we propose a new model for analysing the risk of agitation by utilising the training data obtained from our previous trial. Our key goals are to improve the generalisability of the model and reduce the false-positive rate in the predictions.

Our proposed model uses the information from unlabelled data to understand the data features better and then employ this information in the classification process to improve the generalisability of the model. The model has a self-supervised part that uses different transformations of data and pseudo-labels to train a Convolutional block. The Convolutional block will be used later in the classification training. The proposed classification model is an ensemble model with a Bayesian combination.  


\begin{figure}
    \centering
    \begin{subfigure}{0.5\textwidth}
        \centering
        \includegraphics[width=\linewidth]{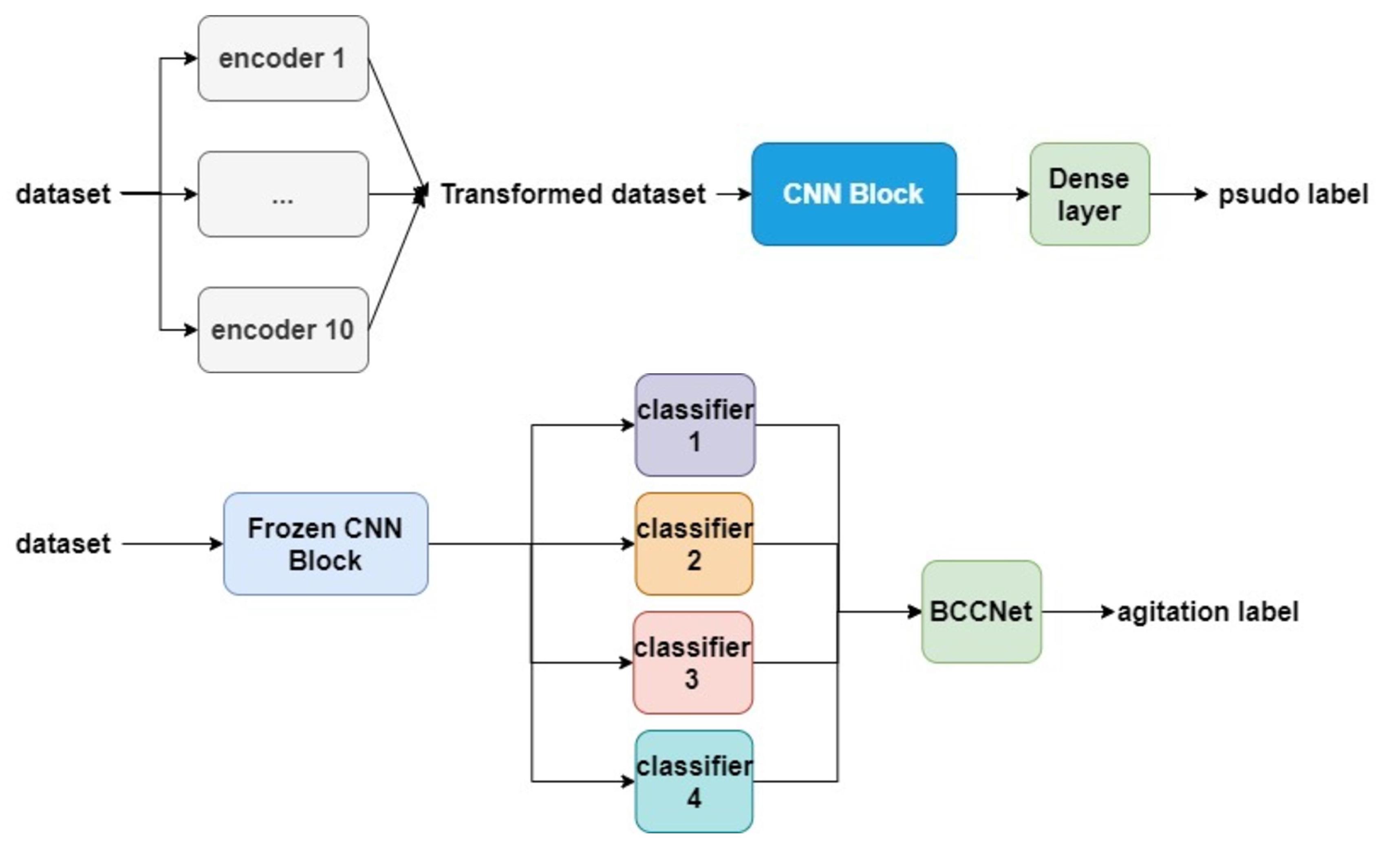}
     \caption{The self-supervised and ensemble agitation classification models.}
     \end{subfigure}
     \begin{subfigure}{0.5\textwidth}
        \centering
        \includegraphics[width=\linewidth]{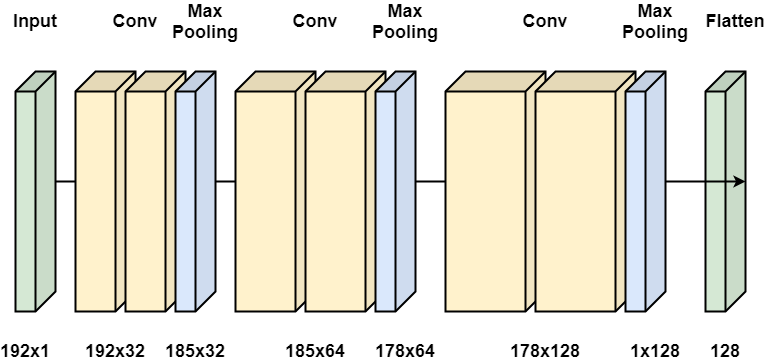}
     \caption{Architecture of the CNN Block which is trained in the first step (self-supervision) to be used in the second step of the proposed model.}
    \end{subfigure}
 \captionof{figure}{An overview of the proposed model. The network uses the self-supervised transformation learning that contains a number of encoders and a CNN block to pseudo label the unlabelled dataset. Model uses the pre-trained CNN block and an ensemble of machine learning models to classify the samples with agitation and not-agitation labels.}
 \label{fig:model_arch}
\end{figure}

\section{Methods}

This section explains how the proposed model uses both unlabelled and labelled data to create a generalisable prediction method for agitation detection. The proposed model has two main parts; a self-supervised transformation and classification model and an ensemble classification model to detect agitation.

\subsection{Self-supervised Transformation Learning}

Our dataset contains a limited amount of labelled data and a vast amount of unlabelled data. We use transformation models to transform the dataset and generate pseudo-labels. This model contains two parts. In the first section, we describe the transformation techniques, and in the second section, we propose the pseudo-label based recognition model. 

\begin{table*}[h]
\centering
\caption{Hyperparameters' values for each encoder.}
\label{tab:encoder}
\resizebox{\linewidth}{!}{%
\begin{tabular}{|c|c|c|c|c|c|c|c|}
\hline
\textbf{Encoder} & \textbf{Layers} & \textbf{No. of latent features} & \textbf{Dropout size} & \textbf{Pooling} & \textbf{Learning rate} & \textbf{Batch size} & \textbf{Optimiser} \\ \hline
1 & 2  & 43 & 0.2 & Max & 0.0001 & 128 & RMSprop\\ \hline
2 & 2  & 43 & 0.4 & Max & 0.003 & 64 & Adam\\ \hline
3 & 3 & 43 & 0.4 & Max & 0.1 & 32 & Adadelta\\ \hline
4 & 1 & 32 & 0.1 & Mean & 0.0001 & 128 & Adam\\ \hline
5 & 1 & 32 & 0.3 & Mean & 0.003 & 32 & Adagrad\\ \hline
6 & 2 & 32 & 0.4 & Mean & 0.01 & 128 & RMSprop\\ \hline
7 & 3 & 32 & 0.4 & Mean & 0.01 & 32 & Adam\\ 
\hline
8 & 1 & 24 & 0.2 & Max & 0.01 & 128 & Adadelta\\ \hline
9 & 3 & 24 & 0.1 & Mean & 0.0005 & 64 & RMSprop\\ \hline
10 & 3 & 24 & 0.1 & Mean & 0.1 & 64 & Adam\\ \hline
\end{tabular}%
}
\end{table*}

\subsubsection{Dataset Transformation and Pseudo Labelling}

We train several under complete convolutional autoencoder architectures with different hyperparameters Table~\ref{tab:encoder}. Autoencoders \cite{autoencoder} can learn how to construct the data by compressing them using an encoder and then reconstructing them using a decoder. We use the encoder sections of the autoencoders as data transformations. If $h$ encodes the data, then $h = f(x)$ where $f$ is the encoder function and $x$ is the input. Similarly, ${x}' = g(h)$ where function $g$ is the decoder and ${x}'$ is the reconstruction of the input data. During the training, the aim is to decrease the Euclidean distance of $x$ and ${x}'$. With convolutional layers \cite{convolutional}, we can detect and capture the temporal and spatial dependencies in the data. Officially, the convolution in the convolutional layer could be written as below:
\begin{equation}
    z^{(L+1)}_{j} = \left (\sum_{k} w_{j} \ast a^{(L)}_{k} \right ) + b_{j}
\end{equation}
where $z^{(L+1)}_{j}$ is the output related to $j^{th}$ neuron in the $(L+1)^th$ hidden layer with inputs from the previous layer. The convolution filter is $w_{j}$ which slides over the previous layer's activation $a^{(L)}_{k}$ where $k$ refers to the index of neuron in the $(L)^th$ layer. $b_{j}$ is the bias term. We used different hyperparameters to construct different encoders. The hyperparameters are the number of convolutional blocks, dropout size, number of extracted features by the encoder, pooling type, optimiser function, learning rate and batch size in the training process. The architectures of encoders are shown in Table \ref{tab:encoder}.

Each convolutional block in an encoder or decoder is a two-dimensional convolutional layer and dropout layer. There is a pooling layer with size $2$ and strides $2$ in each encoder and an upsampling layer with size $2$ in each decoder. The Last layers of the encoder are a flatten layer and a fully connected layer that extract hidden representation or features. Convolutional layers have kernel size $3$, strides $1$ and the padding. Rectified Linear Unit (ReLU) \cite{relu} was chosen as the activation function for all convolutional layers.

\begin{equation}
    ReLU(x) = \begin{cases}
x & \text{ if } x \geq 0 \\ 
0 & \text{ if } x < 0
\end{cases}
\end{equation}

\noindent The Mean Squared Error (MSE) was chosen as the loss function to minimise the reconstruction error in the training of the autoencoders. We use $10$ encoders to transform the dataset, and the corresponding labels for transformations are indicated by $1,...,10$. Label $0$ indicates the original dataset without a transformation.

\subsubsection{Transformation Classification}
We propose a network to classify the transformed datasets. The architecture of the network consists of a CNN block and a Dense block. The CNN block has six convolutional blocks with a max-pooling layer with size $8$, stride $1$, and valid padding after every two blocks. The Dense block has a flatten layer, a fully connected layer with $128$ neurons and another fully connected layer with the number of classes as the number of neurons. The convolutional block includes a one-dimensional convolution layer and a dropout layer with the rate of $0.1$. First, two convolutional blocks have filter size $32$, kernel size $32$, stride $1$, and the same padding and Relu activation function. The following two convolutional blocks have filter size $64$, kernel size $16$, stride $1$, and the same padding and Relu activation function. The last two convolutional blocks have filter size $128$, kernel size $8$, stride $1$, and the same padding and Relu activation function. The CNN block of the network mentioned above can learn the representation of the data in the training process.

\begin{figure}
      \centering
     \includegraphics[width=0.9\linewidth]{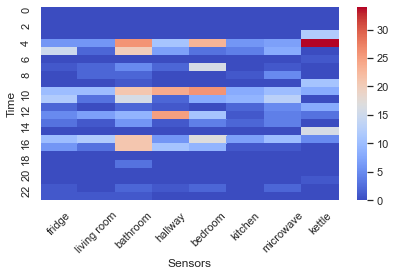}
     \captionof{figure}{A heatmap of $24$ hours of data within a patient's home; the data aggregated within each hour of the day.}
     \label{fig:visual_agg_data}
\end{figure}

\subsection{Ensemble Agitation Classification}
In this stage, we train the model using the true agitation labels. We use the frozen CNN block from the previous step and add agitation classification layers after that. To make a reliable classifier, we use ensemble classification. Ensemble classification includes a number of base classifiers that will be combined to generate the outcome. For the combination part, we use Bayesian fusion (BCCNet \cite{bccnet}). We use $4$ base classifiers which are Naive Bayes \cite{naivebayes}, K-Nearest Neighbour (KNN) \cite{knn}, Support Vector Machine (SVM) \cite{svm} and Gaussian Process (GP) Classifiers \cite{gaussianprocess}. We use these classifiers because each has some advantages. Naive Bayes classifier works well with a small amount of training data, and it is probabilistic. KNN is useful because of simplicity and it has no data assumption. SVM works better with small datasets, and the Gaussian Process classifier is probabilistic and has an integrated feature selection. We combine the results from the classifiers using a neural network and train it with the variational Expected Maximisation (EM)-algorithm. The architecture of the neural network contains a Long Short Term Memory (LSTM) \cite{lstm} layer with $128$ neurons, a fully connected layer with $128$ neurons and Relu activation and L2 regulariser, a flatten layer and a fully connected layer with the number of neurons equal to the number of classes and Sigmoid activation function. We used LSTM in the architecture to learn temporal dependencies in the data.


\begin{table*}[]
\centering
\caption{Baseline models architectures. BCE is the short term for Binary Cross Entropy.}
\label{tab:baseline}
\resizebox{\textwidth}{!}{%
\begin{tabular}{|c|c|c|c|c|c|c|c|}
\hline
\textbf{Model}     & \textbf{Input Layer} & \textbf{Hidden Layers} & \textbf{\begin{tabular}[c]{@{}c@{}}Hidden Layers \\ Activation\end{tabular}} & \textbf{Output Layer} & \textbf{\begin{tabular}[c]{@{}c@{}}Output Layer \\ Activation\end{tabular}} & \textbf{Loss} & \textbf{Optimiser} \\ \hline
\textbf{LSTM}& 24 x 8 & \begin{tabular}[c]{@{}c@{}}LSTM - 200 neurons\\dropout - 0.2\\ LSTM - 200 neurons\\dropout - 0.2\end{tabular} & \begin{tabular}[c]{@{}c@{}}ReLU\\\\ReLU\end{tabular} & 2 & Sigmoid & BCE & Adam (lr=0.0005) \\ \hline
\textbf{BiLSTM} & 24 x 8 & \begin{tabular}[c]{@{}c@{}}BiLSTM - 1000 neurons\\dropout - 0.1\end{tabular} & ReLU & 2 & Sigmoid & BCE & Adam (lr=0.003) \\ \hline
\textbf{VGG} & 8 x 8 x 3 & \begin{tabular}[c]{@{}c@{}}Pre-trained VGG\\ FC - 1000 neurons\\  FC - 1000 neuron\end{tabular} & \begin{tabular}[c]{@{}c@{}}\\ReLU\\ReLU\end{tabular} & 2 & Sigmoid & BCE & Adam (lr=0.0001)   \\ \hline
\textbf{ResNet} & 8 x 8 x 3  & \begin{tabular}[c]{@{}c@{}}Pre-trained ResNet\\ FC - 200 neurons\\  FC - 200 neurons \\ FC - 200 neurons\end{tabular} & \begin{tabular}[c]{@{}c@{}}\\ReLU\\ReLU\\ReLU\end{tabular} & 2 & Sigmoid  & BCE & Adam (lr=0.003) \\ \hline
\textbf{Inception} & 8 x 8 x 3 & \begin{tabular}[c]{@{}c@{}}Pre-trained Inception\\ FC - 200 neurons\end{tabular} & \begin{tabular}[c]{@{}c@{}}\\ReLU\end{tabular} & 2 & Sigmoid & BCE & Adam (lr=0.0001)\\ \hline
\end{tabular}%
}
\end{table*}

\section{Results}

\subsection{Data}

The dataset used in this research has been collected as part of a clinical study regarding people living with dementia from August $2019$ until July $2020$. This clinical study has received ethic approval from South East Coast Surrey NHS Research Ethics Committee (i.e. SURREY, NRES Committee SE Coast (HEALTH RESEARCH AUTHORITY); TIHM REC: 16/LO/1802; IRAS: 211318). There are 96 patients in the study. The dataset contains $14,750$ days of unlabelled data and $417$ days of labelled data. Each participant has been diagnosed with mild to severe dementia and has been stable on dementia medication for at least three months before recruitment. The dataset is a set of continuous environmental sensor data from homes of people with dementia living in the UK. The environmental sensors include PIR, smart power plugs, motion and door sensors. The sensors are placed in the bathroom, hallway, bedroom, living room (or lounge), and kitchen. The sensors are also installed on the fridge door, kettle and microwave (or toaster). A digital platform has been designed to integrate this dataset and help collaboration between clinicians and user groups to support people with dementia \cite{Enshaeifar20}. A clinical monitoring team has been set up to interpret the data daily and review and take action based on the risk alerts generated by the system for different incidents such as; agitation, UTI, abnormal blood pressure, and abnormal body temperature. In this paper, we use agitation alerts. The data label is set as positive if the monitoring team verifies the agitation alert, and a false label is set when they have rejected the alert. We use the environmental data and agitation labels as our dataset. We aggregate the data of sensors within each hour of the day to make the data prepared for the analysis. One example of the data can be seen in Fig. \ref{fig:visual_agg_data}.

\subsection{Evaluation Metrics}

To evaluate our proposed model and compare it with other techniques, we use accuracy, precision, recall, F1-score, area Under the Receiver Operating Characteristic (ROC) curve or AUC metric, and area under the precision-recall curve. These metrics are defined using the terms: True Positive (TP), False Positive (FP), True Negative (TN) and False Negative (FN). TP is the number of samples that have been correctly classified as agitation by the classification method. FP is the number of samples that have been incorrectly classified as agitation. TN is the number of samples that have been correctly defined as not-agitation, and FN is referred to samples that have been incorrectly defined as not-agitation. Accuracy is the measure of how close the classification is to the true classes. In other words, it is the proportion of accurately classified samples to the total number of samples, as shown in equation \ref{eq:accuracy}.

\begin{equation}
\label{eq:accuracy}
\hbox{Accuracy} = \frac{\hbox{TP} + \hbox{TN}}{\hbox{TP} + \hbox{TN} + \hbox{FP} + \hbox{FN}}
\end{equation}

Recall is the measure of classification ability to find all samples truly classified as agitation and defined as the proportion of TP among all the positive samples:

\begin{equation}
\label{eq:recall}
\hbox{Recall} = \frac{\hbox{TP}}{\hbox{TP} + \hbox{FN}}
\end{equation}

Precision is the proportion of TP among all the samples which have been classified as agitated: 

\begin{equation}
\label{eq:precision}
\hbox{Precision} = \frac{\hbox{TP}}{\hbox{TP} + \hbox{FP}}
\end{equation}

F1-score is the weighted average of the precision and recall which is equal to: 

\begin{equation}
\label{eq:f1-score}
\hbox{F1-score} = 2 * \frac{\hbox{Precision} * \hbox{Recall}}{\hbox{Precision} + \hbox{Recall}}
\end{equation}

ROC is a plot of True Positive Rate (TPR) and False Positive Rate (FPR), and AUC measures the area under this curve. TPR and recall are the same, and FPR is shown in equation \ref{eq:fpr}. AUC is the probability of the model ranking a random positive sample higher than a random negative sample.

\begin{equation}
\label{eq:fpr}
\hbox{FPR} = \frac{\hbox{FP}}{\hbox{FP} + \hbox{TN}}
\end{equation}
Precision-recall curve is a plot with y-axis as precision and x-axis as recall which shows the relationship between these two metrics, and as a test metric, we use the area under this plot. This metric is better for testing models on imbalanced data than the area under the ROC curve as the imbalanced dataset with more negative samples makes the ROC plots look better than the model actually performs.

\subsection{The Baseline Models}

We compare the performance of the proposed model with other classification methods such as LSTM network \cite{gers1999learning}, Bidirectional Long Short Term Memory (BiLSTM) network \cite{bilstm}, VGG \cite{vgg}, ResNet \cite{resnet}, Inception \cite{inception}, RF \cite{breiman2001random}, SVM \cite{cortes1995support} and GP classifiers \cite{rasmussen2003gaussian}. The architecture of each model has been optimised using a random search technique. In VGG, Inception, and ResNet classifiers, we use pre-trained models with a classification block afterwards. We note that only the classification blocks have been optimised. Table \ref{tab:baseline} shows the model parameters for LSTM, BiLSTM, VGG, Inception, and Resnet.

RF hyperparameters are as follow; $166$ is considered as the number of estimators, the maximum number of features to consider when looking for the best split is set to the square root of the number of features, maximum depth of the tree is $110$, the minimum number of samples to split the internal node is $2$, minimum samples as a leaf are $1$, and feature bootstrapping is not considered. SVM parameters have been set as follows; regularisation parameter (C) is $100000$, the kernel is Radial basis function (RBF), kernel coefficient is $1e-07$ and no class weight is considered. For GP, the Dot-Product kernel is considered. Other hyperparameters of these models are the default value in the Scikit-learn tool in Python.

\begin{table}[h]
\centering
\caption{Comparison of the proposed model with other models using $10$-fold cross-validation.}
\label{tab:comparison}
\resizebox{\columnwidth}{!}{%
\begin{tabular}{|c|c|c|c|c|c|c|}
\hline
\textbf{Model} & \textbf{Accuracy} & \textbf{Precision} & \textbf{Recall} & \textbf{F1-score} & \textbf{ROC-AUC} & \textbf{PR-REC AUC} \\ \hline
LSTM & 0.58 & 0.27 & 0.26 & 0.26 & 0.49 & 0.32 \\ \hline
BiLSTM & 0.69 & 0 & 0.25 & 0 & 0.56 & 0.51 \\ \hline
VGG & 0.68 & 0.44 & 0.38 & 0.41 & 0.66 & 0.48 \\ \hline
ResNet & 0.69 & 0 & 0.12 & 0 & 0.56 & 0.42 \\ \hline
Inception & 0.7 & 0.47  & 0.35 & 0.38 & 0.67 & 0.45 \\ \hline
Random Forest  & 0.75 & 0.69 & 0.26 & 0.37 & 0.72 & 0.57  \\ \hline
SVM & 0.75 & 0.81 & 0.2  & 0.31 & 0.65 & 0.5 \\ \hline
GP & 0.75 & 0.83 & 0.16 & 0.26 & 0.68 & 0.51 \\ \hline
Proposed Model & 0.63 & 0.58 & 0.59 & 0.59 & 0.59 & 0.51 \\ \hline
\end{tabular}%
}
\end{table}

\subsection{Experiments}

\begin{figure*}
    \centering
    \begin{subfigure}{.3\textwidth}
        \centering
        \includegraphics[width=\linewidth]{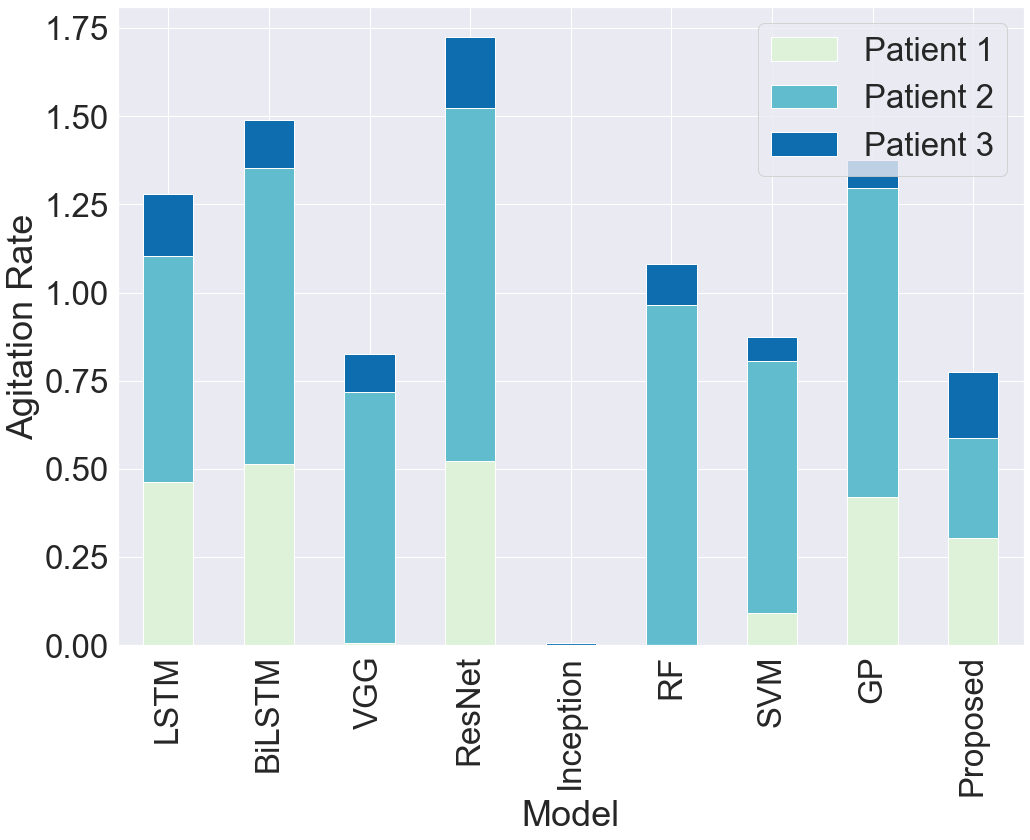}
     \caption{The rate of alerts generated by the models during the testing phase.}
    \end{subfigure}
    \begin{subfigure}{.3\textwidth}
      \centering
     \includegraphics[width=\linewidth]{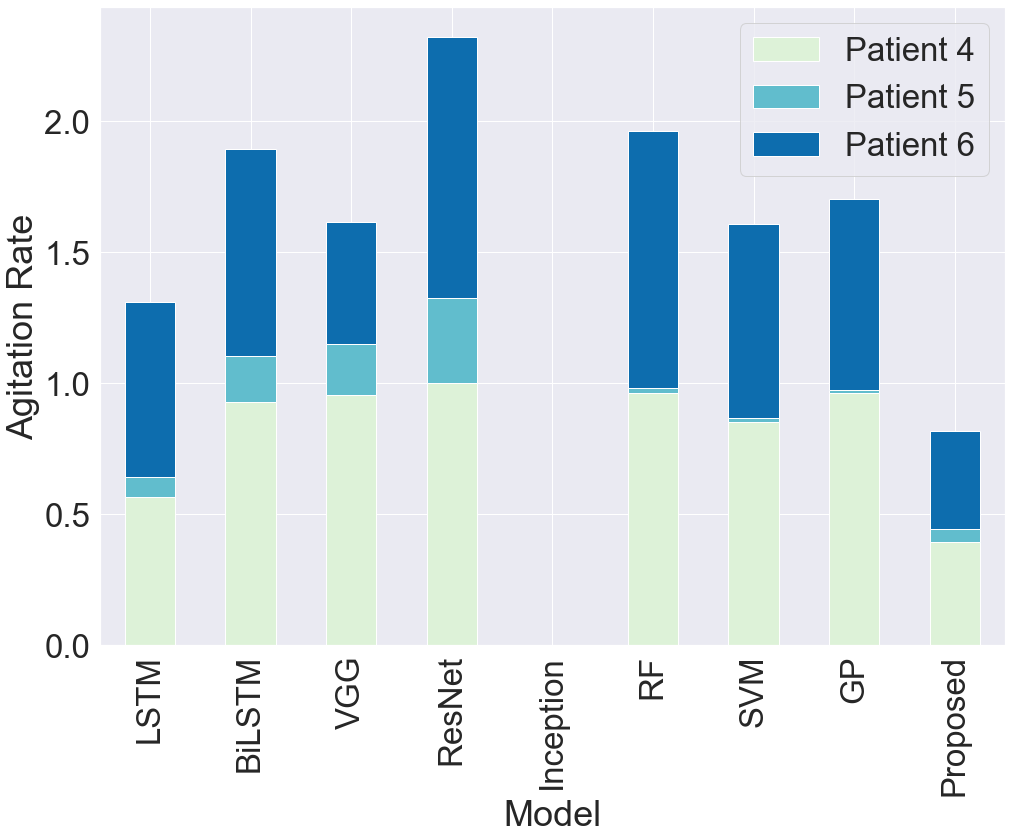}
     \caption{The patients are validated as negative agitation.}
     \label{subfig:b}
    \end{subfigure}
    \begin{subfigure}{.3\textwidth}
      \centering
     \includegraphics[width=\linewidth]{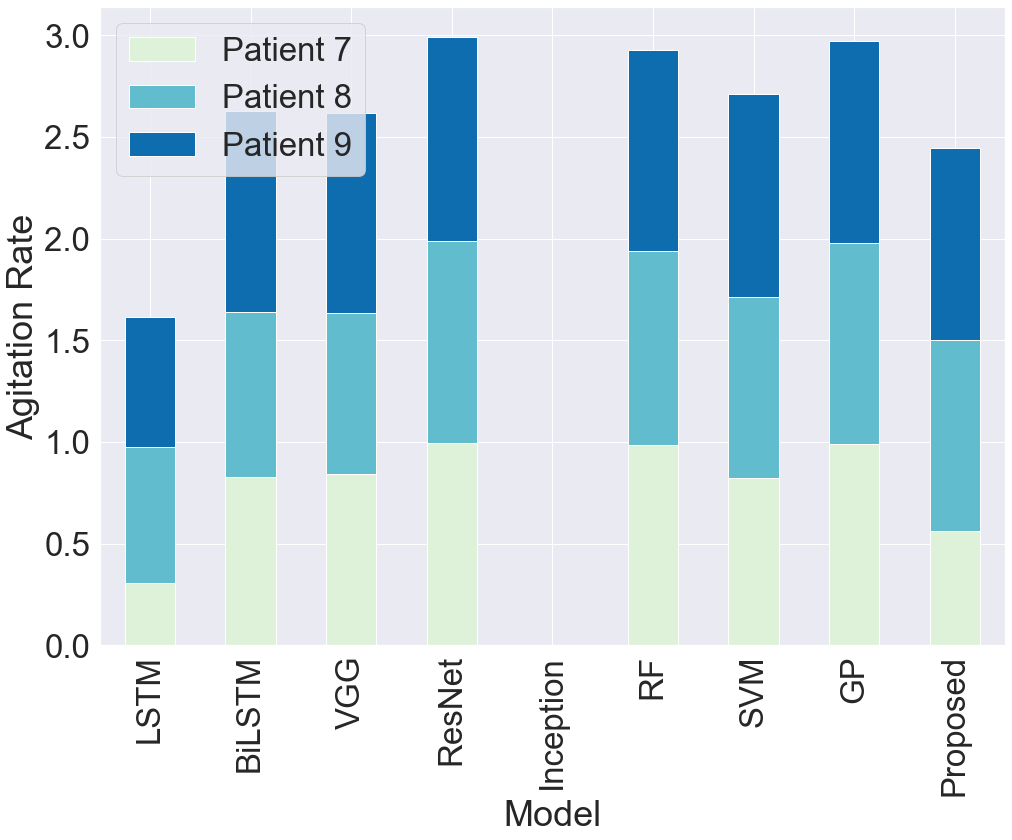}
     \caption{The patients are validated as positive agitation.}
     \label{subfig:c}
    \end{subfigure}
    \caption{Testing the model on a new cohort and with the data that was not seen by the trained model to represent the rate of alerts generated by the models.}
    \label{fig:test_alert_generation}
\end{figure*}

For evaluation, we utilised $10$-fold cross-validation for training and validating our model. In each fold, the data splits into two training and test sets. We report the average of the test results in the cross-validation. The reported results in Table \ref{tab:comparison} are from applying the model to the test set. As shown, the proposed model has a better performance in terms of recall, f1-score and area under the precision-recall curve. As the dataset is imbalanced, it is essential to assess the performance of the model using the metrics that are less affected by class imbalance. The main challenge in imbalance classification evaluation is giving the minority class higher importance. Accuracy is one of the metrics which can be misleading in the imbalanced dataset scenario. In an imbalanced setting, sensitivity or recall can be more interesting than specificity. Moreover, the precision-recall curve can be more useful compared to the ROC curve. Precision and recall are two metrics with a negative correlation that must be balanced as in the f1-score. As shown in Table \ref{tab:comparison}, the proposed model can balance these two better than other baseline models. It can be seen in the results that RF can also perform well. However, as shown in the following subsection, RF cannot perform as good as the proposed model on an independent test set.

\subsection{Generalisation and Discussion}

\begin{figure}
      \centering
     \includegraphics[width=0.9\linewidth]{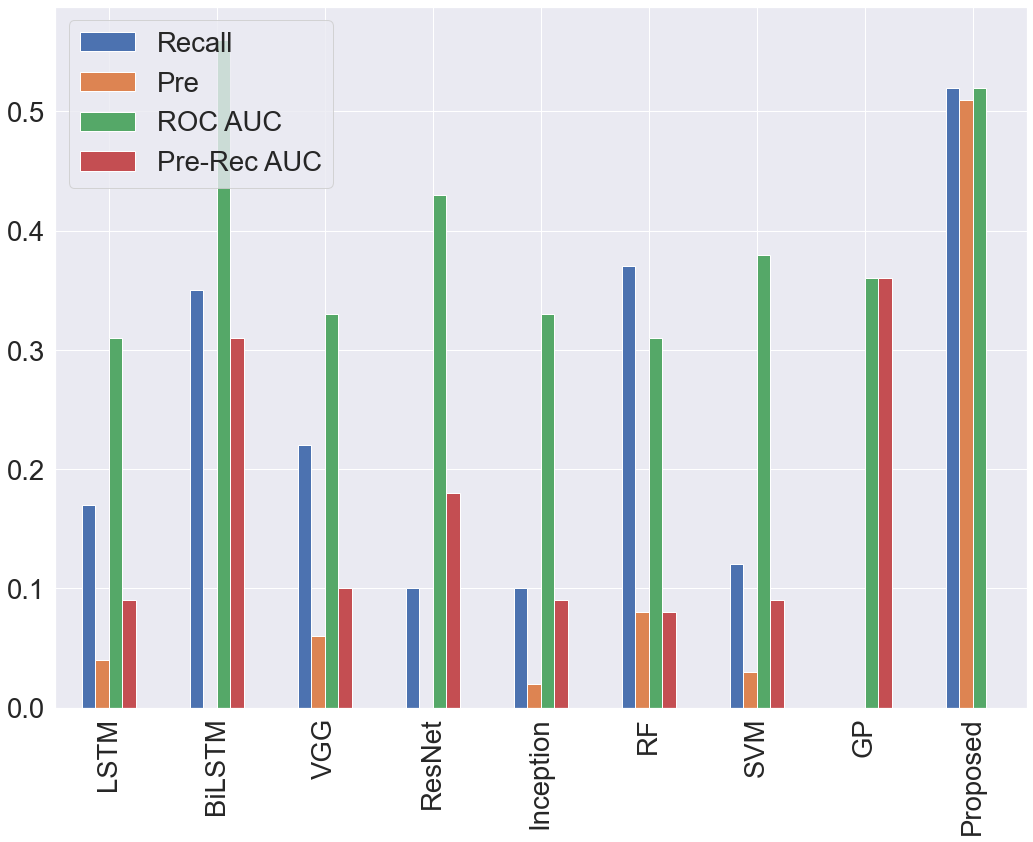}
     \captionof{figure}{Comparison of performance on the hold-out cohort.}
     \label{fig:test_metrics}
\end{figure}

One of the challenges in machine learning for healthcare is the generalisation of the models to the real world. This issue is especially significant in our case when there is not enough training data. We analyse the models' ability in generalisation by testing them on an entirely new group of data that the models did not see. The new dataset contains data from August $2020$ until mid-February $2021$, representing $7,309$ days of data. Fig \ref{fig:test_alert_generation} shows the alerts generated by the models for a number of patients. As can be seen, some models overfit, leading to massive numbers of agitation alerts. The proposed model generally reduces the rate of false alerts. 
In Fig. \ref{subfig:b} and \ref{subfig:c}, we can see the models' generalisability on data from patients with validated cases as not agitation and agitation. The proposed model generates a small number of alerts on not agitation samples and is more sensitive to agitation samples. Then, we tested the performance of the learned models in the previous section on the new hold-out test dataset. As seen in Fig. \ref{fig:test_metrics}, the proposed model outperforms the baseline models.

Semi-supervised models have shown to benefit our setting by learning from extensive unlabelled data and learning to extract features from data collected in an uncontrolled environment. This characteristic helps the model to perform better in a real-world setting and become more generalisable. This can be useful in the healthcare field, which is hard to collect high-quality labelled data.

\section{Conclusion}

In our research, there has been a new form of collecting data for healthcare purposes. This setting, which combines environmental data and machine learning models, can help monitor patients with medical conditions. We proposed a two-stage model. The first stage contains a self-supervised transformation learning model which uses unlabelled data in the learning process. The second stage includes a Bayesian ensemble classification which learns from labelled data. We show that the model outperforms the state-of-the-art models in recall and f1-score values. Another key feature of the model is generalisability which was confirmed by testing the model on a new data cohort.

\section*{Acknowledgment}

This work is supported by Care Research and Technology Centre at the UK Dementia Research Institute. The work is also partially supported by the European Commissions Horizon 2020 (EU H2020) IoTCrawler project (\url{http://iotcrawler.eu/}) under contract number: 779852.

\ifCLASSOPTIONcaptionsoff
  \newpage
\fi

\bibliographystyle{IEEEtran}
\bibliography{IEEEabrv,references}

%

\begin{IEEEbiography}[{\includegraphics[width=1in,height=1.25in,clip,keepaspectratio]{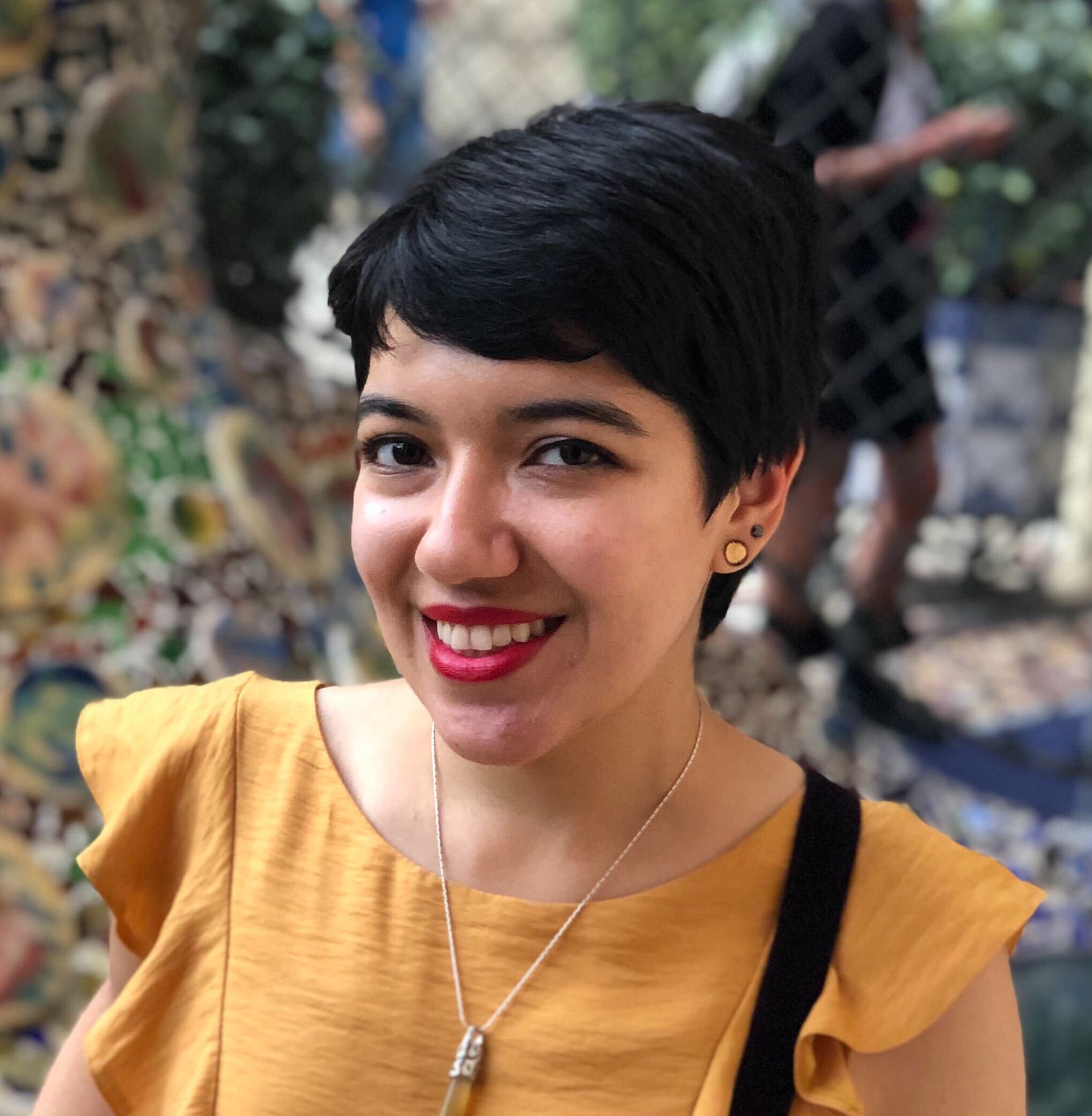}}]{Roonak Rezvani} is currently pursuing her PhD in Machine Learning at the Centre for Vision, Speech and Signal Processing (CVSSP) at the University of Surrey. Her research interests include machine learning in healthcare, big data analytics, time-series data processing and responsible AI.
\end{IEEEbiography}

\begin{IEEEbiography}[{\includegraphics[width=1in,height=1.25in,clip,keepaspectratio]{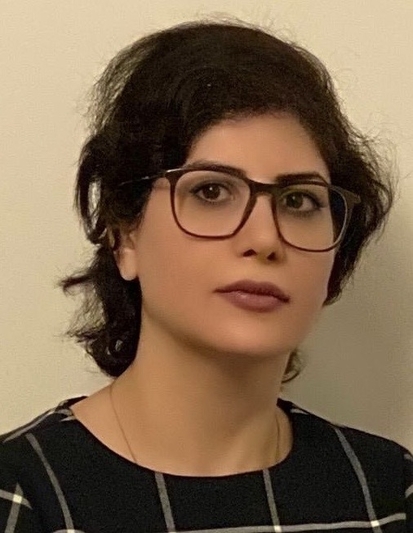}}]{Samaneh Kouchaki}
is Lecturer in Machine learning for Healthcare at the Department of Electronic and Electrical Engineering and a member of the Centre for Vision, Speech and Signal Processing (CVSSP) at the University of Surrey. Her research interests include machine learning, health informatics, biomedical signal processing and computational biology.
\end{IEEEbiography}

\begin{IEEEbiography}[{\includegraphics[width=1in,height=1.25in,clip,keepaspectratio]{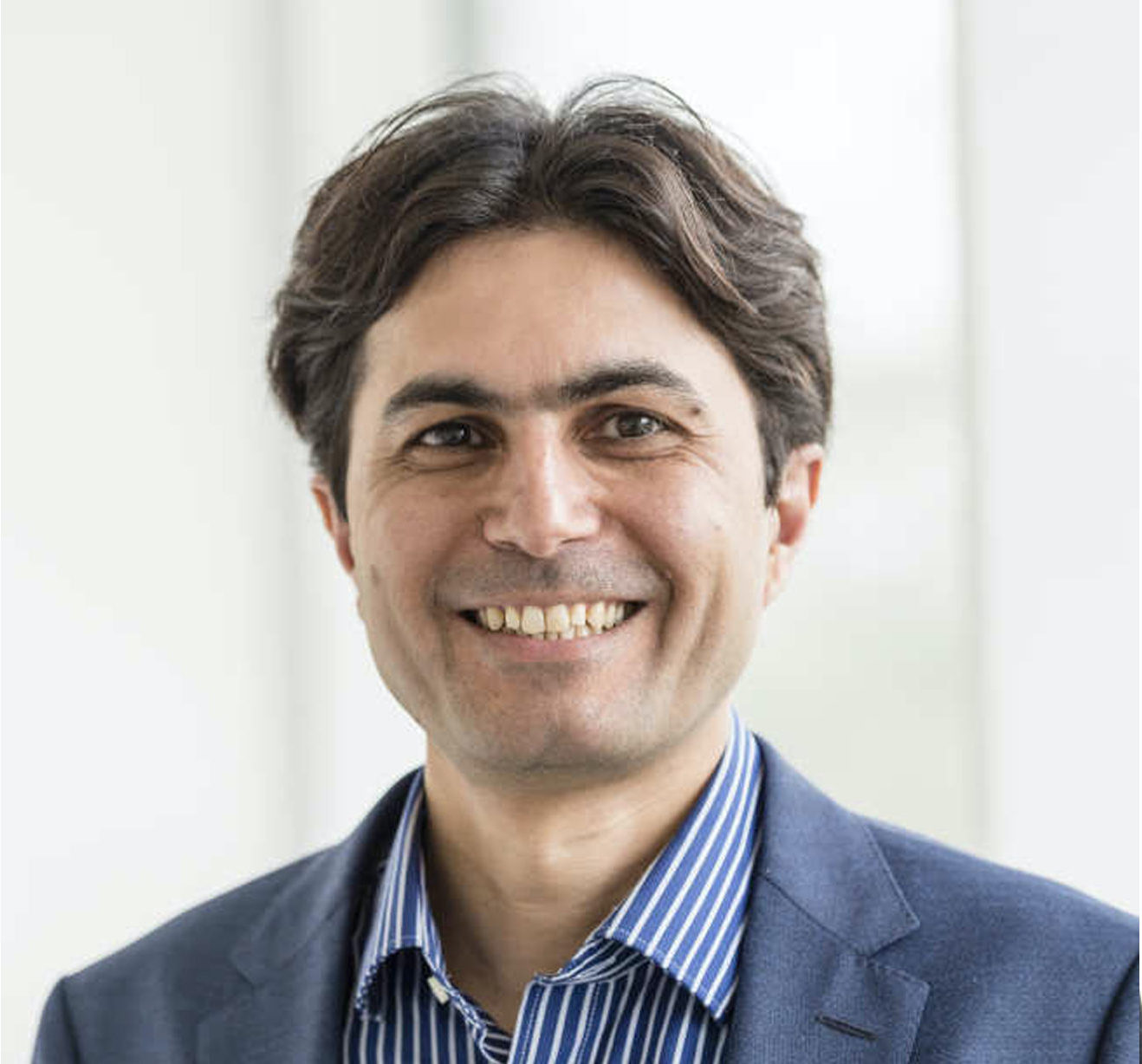}}]{Payam Barnaghi}
is Chair in Machine Intelligence Applied to Medicine in the Department of Brain Sciences at Imperial College London.
He is Co-PI and Deputy Director of the Care Research and Technology Centre at the UK Dementia Research Institute (UK DRI). His research focuses on developing machine learning, adaptive algorithms and Internet of Things solutions for healthcare applications.
\end{IEEEbiography}

\begin{IEEEbiography}[{\includegraphics[width=1in,height=1.25in,clip,keepaspectratio]{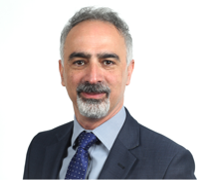}}]{Ramin Nilforooshan} is a Consultant Psychiatrist for older adult psychiatry in Surrey and Borders Partnership NHS Foundation Trust and a Visiting Professor at University of Surrey. He is the CRN Speciality Dementia Lead in Kent, Surrey and Sussex.
\end{IEEEbiography}

\begin{IEEEbiography}[{\includegraphics[width=1in,height=1.25in,clip,keepaspectratio]{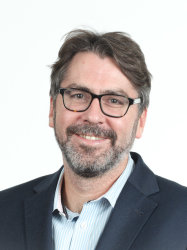}}]{David J. Sharp} is a neurologist and Centre Director of UK Dementia Research Institute (UK DRI) Care Research and Technology, focusing on using technology to enhance the lives of people living with dementia. He is also Scientific Director of the Imperial College Clinical Imaging Facility and Associate Director of the Imperial Centre for Injury Studies.
\end{IEEEbiography}





\end{document}